\documentclass[aps,prl,twocolumn,groupedaddress]{revtex4}

\usepackage{graphicx}% Include figure files
\usepackage{dcolumn}% Align table columns on decimal point
\usepackage{bm}% bold math

\begin{document}

\title{Single Phase Slip Limited Switching Current in 1-Dimensional Superconducting Al Nanowires}% Force line breaks with \\

\author{Peng Li}
\author{Phillip M. Wu}
\author{Yuriy Bomze}
\author{Ivan V. Borzenets}
\author{Gleb Finkelstein}
 \author{A. M. Chang}%
 \email{yingshe@phy.duke.edu}
\affiliation{ Department of Physics, Duke University, Durham NC, 27708}

\date{\today}

\begin{abstract}
An Aluminum nanowire switches from superconducting to normal as the current 
is increased in an upsweep.  The switching current ($I_s$) averaged over  
upsweeps approximately follows the depairing critical current ($I_c$) but 
falls below it. Fluctuations in $I_s$ exhibit three distinct regions of 
behaviors and are non-monotonic in temperature: saturation well below the 
critical temperature $T_c$, an increase as $T^{2/3}$ at intermediate 
temperatures, and a rapid decrease close to $T_c$. Heat dissipation analysis 
indicates that a single phase slip is able to trigger switching at 
low and intermediate temperatures, whereby the $T^{2/3}$ dependence arises 
from the thermal activation of a phase slip, while saturation at low 
temperatures provides striking evidence that the phase slips by macroscopic 
quantum tunneling.

\end{abstract}

\pacs{Valid PACS appear here}% PACS, the Physics and Astronomy
\maketitle

One of the fundamental questions in one-dimensional (1D) superconductivity is 
the nature of the current-induced transition from the superconducting to the 
normal state. Ideally, the maximum current (called the critical current $I_c$) 
is set by the depairing mechanism, where the Cooper pairs are destroyed by 
the electron velocity. Experimentally, however, the depairing limit is difficult 
to achieve. The maximum current is often limited by either the self-generated 
field in 3D samples or the motion of magnetic flux 2D thin film (2D) 
\cite{TinkhamBook}. It was believed depairing $I_c$ may be achievable in a 
narrow superconducting wire. Nevertheless, fluctuation effects, specifically 
the spatio-temporal fluctuations of the order parameter known as phase slips 
(explained below), can induce premature switching \cite{HysteresisIV,Goldbart}. 
A recent study reported phase-slip-induced switching \cite{NaturePhysics}. 
Unexpectedly, it was found that the switching current ($I_s$)fluctuations 
increased {\it monotonically} with decreasing temperature.  To date, 
clear evidence is lacking as to whether a single phase slip is capable of 
inducing switching . Establishing the relationship between individual phase slips 
and switching provides a tool to study the phase slip, to help establish whether 
they are caused by thermal fluctuations or by macroscopic quantum tunneling 
\cite{Little,LA,MH,Giordano.PRL,Nature,Lau.PRL,Highfield.B,Fabio.PRL}. 
Such studies not only elucidate the fundamental physics, e.g. a 
superconductor-insulator transition caused by QPS 
\cite{Nature,QPStheory.Zaikin1,sergei}, but also provide 
the basis for applications from a new current standard to 
quantum qubits \cite{nanowire.qbit,CurrentJJ}.

\begin{figure}
\includegraphics[width=0.5\textwidth]{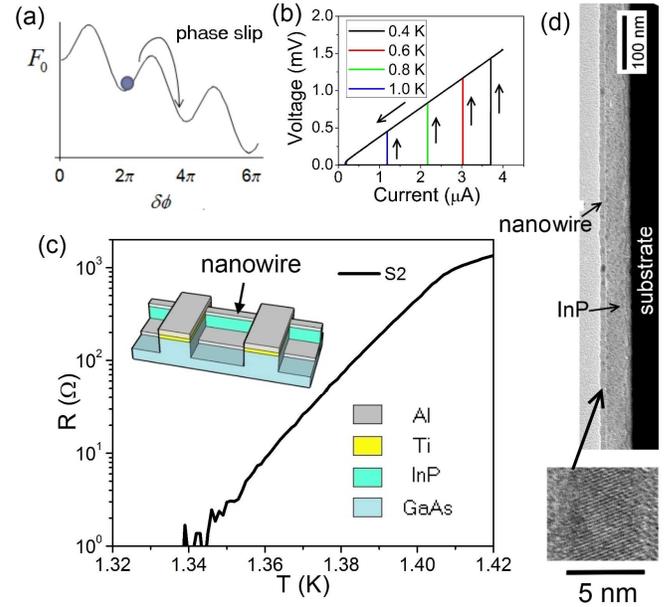}
\caption{\label{fig:fig1} (a) Schematic of a washboard potential. (b) Hysteretic IV curves and 
(c) resistive transition versus temperature for nanowire S2.  (d) Transmission Electron Microscope images of 
a typical nanowire. The long and uniform nanowire sits on an InP ridge.}
\end{figure}

The phase slip is a topological event, during which the 
superconducting order parameter phase between two adjacent regions of the 
superconductor changes by 2$\pi$ over a spatial distance of the order of
the coherence length.  
%, with a core which is normal, for at least one point in spatial position.  
To understand the destruction of supercurrent in a nanowire, one may gain valuable 
insight from phase slips in a Josephson junction \cite {TinkhamBook}.  
In both systems, the motion of the phase is described by a tilted washboard 
potential (Fig. 1a).  Josephson junctions are classified within a Resistively and 
Capacitively Shunted Junction (RCSJ) model as either under- or over-damped, 
depending on whether the quality factor, $Q=\sqrt{2eI_cC/\hbar}R$, 
is greater or less than 1.  An under-damped junction is readily driven 
normal by a single phase slip event; the phase keeps running downhill subsequent 
to overcoming the free-energy barrier, as damping is insufficient to 
retrap into a local minimum.  Thus, the junction exhibits zero resistance 
up to $I_s$ and the voltage is hysteretic in an current-voltage (IV) 
measurement.  In an over-damped junction, the phase moves diffusively 
between minima;  the IV is nonlinear and hysteresis is often not present.  The 
different situations thus lead to differing characteristics. 

Due to an extremely small capacitance, a nanowire is believed to be 
heavily over-damped ($Q\ll1$).  
Recently, experimental evidence has accumulated
indicating that heating can also lead to hysteretic behavior in overdamped 
Superconducting-Normal-Superconducting bridges\cite{Hysteresis.SNS}.  
Here, we report on measuring the fluctuations of $I_s$ 
in Al superconducting nanowires. In stark contrast to the previously 
reported monotonic increase with decreasing temperature \cite{NaturePhysics}, 
the fluctuations in our Al nanowires is non-monotonic with three distinct 
regions of behaviors. Below $\sim$ 0.3 $T_c$, a clear saturation of the fluctuations 
is observed indicating the switching to be caused by a quantum-phase-slip 
(QPS). At intermediate temperatures $\sim 0.3 - 0.6 T_c$, 
the fluctuations increase as $T^{2/3}$, signifying 
that switching is caused by a thermal phase slip (TAPS).  
At high temperatures above $\sim$ 0.6 $T_c$, the rapid decrease of fluctuations 
points to multiple TAPSs triggering the switching.  Although in appearance, this behavior 
is reminiscent of an under-damped Josephson junction, for $T < 0.6 T_c$, 
quantitative estimation demonstrates that heat generated by a single phase slip 
likely causes a thermal runaway, triggering switching \cite{Goldbart}.  
Because a single QPS or TAPS is sufficient to trigger switching close to $I_s$, the 
resistance of the SC state at current levels below $I_s$ can remain zero 
much of the time, with occasional jumps as a rare single phase slip event occurs.   

\begin{figure}
\includegraphics[width=0.5\textwidth]{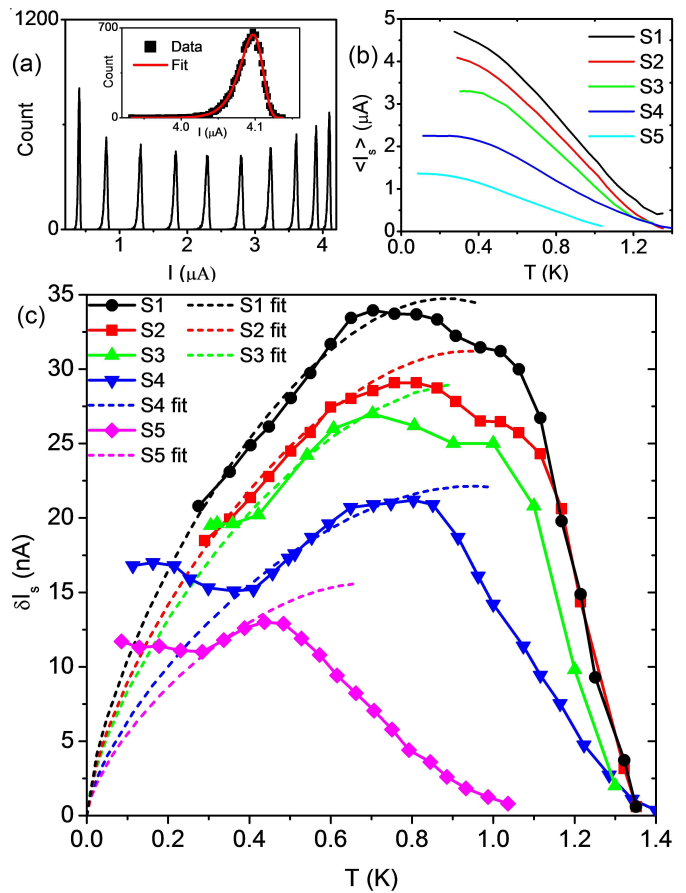}
\caption{\label{fig:fig2} (a) $I_s$ distribution for S2 at different temperatures--right to left: 
0.3 K to 1.2 K in 0.1 K increments. Inset shows the $0.3 K$ distribution, fitted by 
the Gumbel distribution \cite{Gumbel}. (b) $\langle I_s \rangle$ versus temperature. 
(c) Symbols--$\delta I_s$ versus temperature. Dashed lines--fittings 
in the single TAPS regime using Eq. \ref{eq:twothird}.  An additional scale factor  
of 1.25, 1.11, 1.14, 0.98 and 1.0, for S1 - S5, respectively (average $1.1 \pm 0.1$), 
is multiplied to match the data.  Alternatively, a $\sim 6 \%$ adjustment in the exponent 
fits the data without the scale factor.}
\end{figure}

\begin{table}
\caption{\label{tab:specs} Specifics of the samples.}
\begin{ruledtabular}
\begin{tabular}{  l  c  c  c  c  c  }
Sample & S1 & S2 & S3 & S4 & S5\\ \hline \hline
Length ($\mu$m) & 1.5 & 10 & 10 & 10 & 3 \\ \hline
Width (nm) & 10.0 & 9.3 & 8.4 & 7.0 & 5.4 \\ \hline
R/L \footnotemark (K$\Omega/\mu m$) & 0.33 & 0.38 & 0.50 & 0.82 & 1.17 \\ \hline
$I_m$ \footnotemark ($\mu$A) & 4.7 & 4.1 & 3.3 & 2.3 & 1.4 \\ \hline
$T_s$ \footnotemark (K) & 1.36 & 1.33 & 1.33 & 1.25 & 1.03 
\footnotetext[1]{Resistance per unit length of the nanowires in normal state.}
\footnotetext[2]{Maximum switching current has measured}
\footnotetext[3]{Transition temperature, below which the nanowires show zero measured resistance}
\end{tabular}
\end{ruledtabular}
\end{table}

Five nanowires were studied (TABLE \ref{tab:specs}).
Each end of a nanowire is connected to a large 2D superconducting pad
rather than to a normal metal pad \cite{Fabio.PRL}. 
S3 was obtained by further oxidizing the surface of S2.  
The fabrication was described previously \cite{Fabio.APL}. 
The superconducting coherence length at base temperature $\xi_0 \sim 
100 nm$. The length of the wires range from 15$\xi_0$ to 100$\xi_0$, 
while the width is roughly 1/10$\xi_0$. The resistivity (4.5 $\mu\Omega$cm;  
same as co-evaporated films), along with the inverse proportionality 
to 20\% accuracy between normal resistance per-unit-length 
and $I_s$ at base temperature, indicates that there are no 
resistive tunneling barriers. These nanowires are in the fully metallic limit, 
($k_F l \sim 60 >> 1$, where $k_F$ is the Fermi wavenumber, and $l$ 
the electron mean-free-path), in contrast to those studied 
by Sahu {\it et al.} \cite{NaturePhysics} which appear to be grainy with 
$k_F l$ is much closer to 1, and for which Coulomb effects may be important.

S1 and S2 were measured in a $^3$He system and S3, S4 and S5 in a 
dilution refrigerator.  To fully remove interference from unwanted noise, 
each electrical line is equipped with RL filters (1 MHz cutoff) 
at room temperature, Thermocoax cables (1 GHz cutoff) extending 
to the mixing chamber of the dilution refrigerator, 
and RC filters (34 KHz cutoff) at the mixing chamber.  For the current 
sweep a sawtooth waveform was used at a repetition of 10 Hz.  
The upsweep ramp rate was 50 $\mu$A/s for S1, S2 and S3 and 25 $\mu$A/s for S4 and S5 .  
Decreasing the rate by a factor of 10 yielded nearly identical results.  
Immediately after a voltage jump, the current was 
turned off, reducing the resistive heating time in the normal state to less than 
100 $\mu$s, and ensuring adequate time to re-thermalize the sample ($\sim$ 10$^{-7}$s) 
before the next cycle.

At vanishing current, he nanowires become superconducting below the switching 
temperature $T_s(I \rightarrow 0)$. In Fig. 1(c), the resistive transition is broadened 
due to TAPSs. To measure $I_s$ fluctuations, 
we performed $\sim$ 10,000 I-V sweeps at each temperature, recorded the upsweep $I_s$ 
and plotted the histogram (Fig. 2(a)). The probability density 
function $P(I)$ obeys the expression (suppressing the subscript in $I_s$): 
\begin{equation}
P(I)= \Gamma (I) \left (dI/dt \right )^{-1} \left ( 1-\int_{0}^{I}P(u)du \right )
\label{eq:PI}
\end{equation}
\noindent
where $dI/dt$ is the ramping rate, and $\Gamma(I)$ is the switching 
rate at current $I$ \cite{Fulton}.  If a single phase slip triggers switching, 
the switching and the phase slip rates are identical, enabling to extract 
the phase slip rate from the distribution.  Using the single TAPS rate 
$\Gamma(I) \sim \exp(F(T,I)/k_BT)$, where $F(T,I)$ is free energy barrier, 
and linearizing the current dependence of $F(T,I)$, the solution for $P(I)$ 
is in the form of a Gumbel distribution \cite{Gumbel}.  
An example of the fitting to this functional form to data is shown in the 
inset to Fig. 2(a).  For each distribution, we deduce the mean value 
$\langle I_s \rangle$ and the standard deviation $\delta I_s$, as shown 
in Figs. 2(b) and (c), respectively.

In the single TAPS regime, the fluctuation in $I_s$ is approximately proportional to 
\begin{equation}
\delta I_s \sim (k_B T/\phi_0)^{2/3} I_c(T)^{1/3}
\label{eq:twothird}
\end{equation}
\noindent
where $I_c(T)$ is the depairing $I_c$ at temperature $T$. The $T^{2/3}$ is from the exponent in the current 
dependence of $F(T,I)$, $F(T,I)=F(T) (1-I/I_c)^{3/2}$  \cite{MQT.Clarke}.  
In all samples, $\delta I_s$ in the intermediate temperature range 
($\sim$ 0.3$T_c$ - 0.6$T_c$) can be fitted by Eq. \ref{eq:twothird} 
very well, shown in Fig. \ref{fig:fig2}(c). The good agreement between 
data and theoretical fittings indicates that switching is 
induced by a single TAPS. 

When the nanowire has narrower width, it becomes more probable for the 
phase to undergo a macroscopic quantum tunneling process through the 
barrier.  The QPS rate is proportional to $\exp(-\alpha F(T,I)/\Delta)$, where 
$\alpha$ is constant of order unity,  and different possibilities for $\Delta$ 
have been proposed.  These include $\hbar/\tau_{GL}$, with $\tau_{GL}$  
the Ginzburg-Landau time, and the superconducting gap 
\cite{Giordano.PRL,NaturePhysics,QPStheory.Zaikin2}. 
In S4 and S5, we find a slight increase of the fluctuations with decreasing 
temperature in the QPS regime (below $\sim$ 0.3$T_c$), consistent  
with $\Delta$ scaling as the superconducting gap.

\begin{figure}
\includegraphics[width=0.45\textwidth]{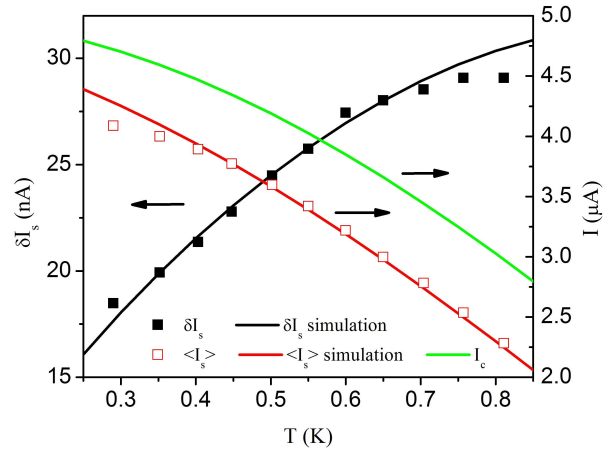}
\caption{\label{fig:fig3} $\delta I_s$ and $\langle I_s \rangle$ fitted 
by TAPS model:  Squares--experimental data; solid lines--simulation.  
$I_c(0)$=5 $\mu$A and $T_c$=1.5 K are two fitting parameters.  
A constant scaling factor $\sim 1.15$ is multiplied to the calculated 
$\delta I_s$ to match the data.  }
\end{figure}

To check the approximate linearized expressions for 
$\delta I_s$ and $\langle I_s \rangle$ in the single TAPS regime, 
a full numerical simulation is performed to solve Eq. \ref{eq:PI}, 
as shown in Fig. \ref{fig:fig3}. The TAPS rate is given by: 
\begin{equation}
\Gamma_{TAPS} (T,I)=\Omega_{TAPS} \exp\left(\frac{-F(T,I)}{k_B T}\right)
\label{eq:Gamma}
\end{equation}
\noindent
where $\Omega_{TAPS}=L/\xi_T \sqrt{F(T,I)/k_B T}/\tau_{GL}$ is 
the attempt frequency, $\tau_{GL}=\pi \hbar/(8k(T_c-T))$, 
$\xi_T$ is the superconducting coherence length at temperature $T$, 
and $F(T,I)=\sqrt{6} \hbar/(2e) I_c(T) (1-I/I_c(T))^{3/2}$ is the 
free energy barrier \cite{NaturePhysics}.  The zero current free 
energy barrier, $F(T)=\sqrt{6} \hbar I_c(T)/2e$, bears 
similarity to the Josephson energy $E_J=\hbar I_c/2e$ \cite{HysteresisIV}. 
To extend to the entire temperature range, a phenomenological 
$I_c(T) = I_c(0)(1 - (T/T_c)^2)^{3/2}$ was employed \cite{Bardeen}.  
Good agreement is achieved up to 0.8 K ($\sim$ 0.6$T_c$) in S2 as 
shown in Fig. \ref{fig:fig3}.  Other samples exhibit a similar agreement.

Above 0.8 K, $\delta I_s$ falls below the simulated value, at 
first decreasing gradually, then rapidly.  This behavior is associated with  
the need for more-than-one phase slips to heat up the wire as the 
current drops \cite{Goldbart}.  A similar decrease is familiar in 
Josephson junctions within the phase diffusion regime, where multiple 
phase slips are required to induce switching \cite{Collapse1,Collapse2}.

To achieve a consistent picture, it must be demonstrated that heat generated 
by a single phase slip is sufficient to raise the local temperature 
and trigger switching.  A single phase slip deposits an energy 
$\phi_0I$ in a time $\phi_0I/(I^2R_{core}) \sim 50 ps$, where 
$\phi_0 = h/2e$ is the flux quantum, and $R_{core}$ is the normal state 
resistance of the phase slip core.   Due to the low $R_{core}$ 
($\sim 100 \Omega$), this energy is deposited predominantly in the 
normal core rather than removed via plasmon emission \cite{sergei}.  
Heat loss through the InP ridge is also ineffective.  
Immediately after the phase slip, the hot normal electrons are decoupled 
from the superconducting electrons for a duration the charge-imbalance time 
$\tau_{imb}$.  Within this time, heat diffuses out primarily within the 
normal electron component to a charge imbalance distance, 
$\Lambda_{imb} \sim \sqrt{D \tau_{ imb}}$, where $D$ is diffusion 
coefficient, before it can be transferred to the superconducting electrons 
to raise their effective temperature to $T_f$.\cite{Goldbart}  
Following transfer, this entire region of size $2 \times \Lambda_{imb}$ either 
becomes normal or returns to superconducting depending on whether $I_c$ 
at the elevated electronic $T_f$ is exceeded or not \cite{footnote}.  
If exceeded, this region becomes normal; its resistance contributes 
further to heating, causing a thermal runaway.  $T_f$ can be estimated as
\begin{equation}
T_f=\sqrt{T_0^2+\phi_0 I/(\gamma A \Lambda_{imb})}
\label{eq:heating}
\end{equation}
\noindent
where $\gamma=C_v/T=135 J/(m^3K^2)$, $C_v$ is the specific heat of normal Al \cite{Pobell}, 
A is cross section area, and $T_0$ is the ambient temperature. 
Setting $\Lambda \sim 0.8 \mu m$, Fig. 4 shows that a boundary 
between the single TAPS regime and the multiple TAPS regime occurs around 
0.7 K ($\sim$ 0.3$T_c$), consistent with our experimental result. This value 
for $\Lambda_{imb}$ is close to the findings in recent experiments on Al 
wires of sub-$\mu m$ diameter \cite{CIB1}, and is expected to be temperature 
independent \cite{Klapwijk.APL}.  The same analysis yields a boundary 
of 0.7 K, 0.6 K, 0.6, K and 0.45 K for S1, S3, S4, and S5, respectively, 
consistent with the behavior in Fig. 2(c). 

\begin{figure}
\includegraphics[width=0.4\textwidth]{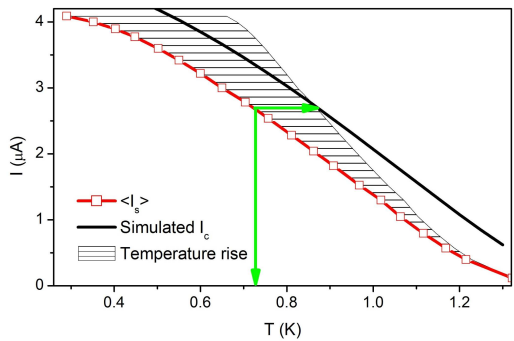}
\caption{\label{fig:fig4} Temperature boundary between single- and 
multiple-phase-slip regimes. The red curve is measured $\langle I_s \rangle$. 
The thick black curve is simulated depairing $I_c$. Following the black shaded 
region toward the right, the thin black curve shows the electronic temperature rises 
due to a single phase slip event at current $I$ according to Eq. \ref{eq:heating}. 
Where the thick black curve exits the shaded area marks a change from 
single- to multiple-phase-slip regimes. The corresponding ambient 
temperature boundary ~ 0.7 K can be found by moving horizontally to the left 
to $\langle I_s \rangle$ (shown by green arrows).}
\end{figure}

In summary, we demonstrate that 1D Al superconducting nanowires can be switched into 
the normal state by a single phase slip, over a sizable temperature range.  At low $T$, 
QPS-induced switching was found in the narrower wires.   In the single TAPS 
regime, $I_s$ fluctuations are proportional to $T^{2/3}$.  The fluctuations decrease 
at higher temperature, where multiple phase slips are needed to trigger switching.  
Heating by phase slips appears to play a major role in the switching process.  
The behavior found is likely relevant to nanowires of different materials, 
due to the commonality of restricted geometry.  

\begin{acknowledgments}
We thank Sergei Khlebnikov for useful discussions. The project is supported 
by NSF DMR-0701948 and Institute of Physics, Academia Sinica, Taipei. 
\end{acknowledgments}

%\bibliography{PengPRL051211.bib}
%\bibliographystyle{apsrev}

\end{document}